\documentclass[aps,prl,twocolumn,superscriptaddress,showkeys,preprintnumbers]{revtex4-1}

\usepackage{amsmath,amssymb,ifpdf,url
             ,mathrsfs,slashed,multirow,tabularx,type1cm}
\usepackage{graphicx,color}

\ifpdf        
\usepackage{graphics}
\else                          
\usepackage[dvipdfmx]{graphics}      
\fi

\allowdisplaybreaks

\definecolor{BlueViolet}{rgb}{0.2, 0.00, 0.7}
\definecolor{Blue}{rgb}{0.15, 0.00, 0.9}
\usepackage[
colorlinks=true,linkcolor=Blue,citecolor=Blue,
urlcolor=BlueViolet]{hyperref}

\newcommand{\GeV}{\,{\rm GeV}}

\newcommand{\Slash}[1]{{\ooalign{\hfil \hspace*{-5pt}~#1\hfil\crcr\raise.167ex\hbox{/}}}}
\def\be{\begin{equation}}
\def\ee{\end{equation}}

\def\({\left(}
\def\){\right)}
\def\<{\langle}
\def\>{\rangle}
\newcommand{\non}{\nonumber \\ }
\newcommand{\matl}{\left( \begin{array}}
\newcommand{\matr}{\end{array} \right)}

\def\beq#1\eeq{\begin{align}#1\end{align}}

\def\GeV{\text{GeV}}

\usepackage{accents}
\newcommand\brobor{\smash[b]{\raisebox{0.6\height}{\scalebox{0.5}{\tiny(}}{\mkern-1.5mu\scriptstyle-\mkern-1.5mu}\raisebox{0.6\height}{\scalebox{0.5}{\tiny)}}}}

\newcommand{\real}{\textrm{Re}\,}
\newcommand{\imag}{\textrm{Im}\,}

\begin{document}

\preprint{TTP17--031}

\title{\boldmath Direct \textit{CP} Violation in $K \to \mu^+ \mu^-$ \unboldmath}

\author{Giancarlo D'Ambrosio}
\email{gdambros@na.infn.it}
\affiliation{INFN-Sezione di Napoli, Via Cintia, 80126 Napoli, Italia}

\author{Teppei Kitahara} \email{teppei.kitahara@kit.edu} 
\affiliation{Institute for Theoretical Particle Physics (TTP), Karlsruhe Institute of Technology, Engesserstra{\ss}e 7, D-76128 Karlsruhe, Germany}
\affiliation{Institute for Nuclear Physics (IKP), Karlsruhe Institute of
Technology, Hermann-von-Helmholtz-Platz 1, D-76344
Eggenstein-Leopoldshafen, Germany}

\date{\today}

\begin{abstract}
A rare decay $K_L \to \mu^+ \mu^- $ has been measured precisely, while a rare decay $K_S \to \mu^+ \mu^- $ will be observed by an upgrade of the LHCb experiment.
Although both processes are almost \textit{CP}-conserving decays, 
we point out that an interference contribution between $K_L$ and $K_S$ in the kaon beam emerges from a  genuine direct \textit{CP} violation.
It is found that the interference contribution can change $K_S \to \mu^+ \mu^-$ standard-model predictions at $\mathcal{O}(60\%)$.
 We also stress that  an unknown sign of $\mathcal{A}(K_L \to \gamma \gamma)$ can be determined by a measurement of the interference, which 
can much reduce a theoretical uncertainty of $\mathcal{B}(K_L \to \mu^+ \mu^-)$.
We also investigate the interference in a new physics model, where 
 the $\epsilon'_K / \epsilon_K$ tension is explained by an additional $Z$-penguin contribution.
\end{abstract}

\maketitle 
\vspace{-0.5cm}


Rare kaon decays  have played a crucial role in flavor physics; now this  physics program is even more exciting due to the 
NA62 experiment at CERN, which aims  to reach a precision of 10 \% in $\mathcal{B}(K^+ \to \pi ^+\nu  \overline{\nu}) $ compared to the standard model (SM) in 2018 \cite{Rinella:2014wfa,Romano:2014xda},  and the KOTO experiment  at  J-PARC, which aims, as a  first  step,   at  measuring $\mathcal{B}(K_L \to \pi ^0 \nu  \overline{\nu} )$ around the SM sensitivity \cite{Komatsubara:2012pn,Shiomi:2014sfa,KOTO}. The LHCb experiment also has an impressive kaon physics program \cite{LHCBK}.
 New physics motivated  from   the $\epsilon'_K / \epsilon_K$ tension  \cite{Bai:2015nea,Buras:2015yba, Kitahara:2016nld} or $B$-physics anomalies may be tested in rare kaon decays  too.
Experimentally kaons in  two muons in the final state can be considered gold channels, and  this motivates theoretical studies.
 
Within the SM, the branching ratios are predicted to be \cite{Ecker:1991ru, Isidori:2003ts, Gorbahn:2006bm}
\begin{align}
\mathcal{B}(K_L \to \mu^+ \mu^-)_{\rm SM} &=  
 \begin{cases}
 \left(6.85 \pm 0.80\pm 0.06\right) \times 10^{-9} (+),\\
  \left( 8.11 \pm 1.49 \pm 0.13\right) \times 10^{-9} (-),
  \end{cases}\label{eq:KLmumu:SM} \\ 
\mathcal{B}(K_S \to \mu^+ \mu^-)_{\rm SM} &= \left( 4.99 \,({\rm LD}) + 0.19 \,({\rm SD}) \right) \times 10^{-12} \non
&=\left(5.18  \pm 1.50\pm 0.02 \right) \times 10^{-12},
\label{eq:KSmumu:SM}
\end{align}
where the first uncertainty comes from long-distance contributions and the second one denotes remaining theoretical uncertainties including the Cabibbo-Kobayashi-Maskawa (CKM) parameters.
The long-distance (short-distance) contribution to $\mathcal{B}(K_S \to \mu^+ \mu^-)_{\rm SM} $ is indicated by LD (SD). 
Here,  the leading chiral contribution at ${\cal O} (p^4)$,  $K_S\to\pi^+ \pi ^- \to \gamma\gamma \to {\mu^+ \mu^-}$,  is theoretically clean \cite{Ecker:1991ru, footnote:K+loop}: $K_S\to\pi^+ \pi ^-$ is described 
in terms of 
$G_8$ (and $G_{27}$) which 
 represents the leading coupling of the $| \Delta S| = 1$ nonleptonic weak Lagrangian \cite{DAmbrosio:1994fgc}
and $\textrm{sgn}(G_8) < 0$ is taken; we just assume the sign predicted by the $| \Delta S|  = 1$ partonic Lagrangian computing the hadronic matrix elements of four-quark operators in the large-$N_C$ limit (or employing naive factorization)  \cite{Isidori:2003ts,Pich:1995qp, Gerard:2005yk}. 
The values of Eqs.\,(\ref{eq:KLmumu:SM}) and (\ref{eq:KSmumu:SM}) are based on the best-fit result for the CKM parameters  
 in Ref.\,\cite{ckmfitter}.
 One should note that $\mathcal{B}(K_L \to \mu^+ \mu^-)_{\rm SM}$ depends on
 an unknown  sign of  $\mathcal{A}(K_L \to \gamma \gamma)$. 
 Indeed, differently from the previously discussed $K_S$ decay,  the leading ${\cal O} (p^4)$  {of}
$\mathcal{A}(K_L \to \gamma \gamma  \to  \mu^+ \mu^- )$  given by the Wess-Zumino anomaly \cite{Wess:1971yu} is vanishing due to the delicate cancellation enforced by the Gell-Mann--Okubo formula of  the two contributions  with $\pi^0$ and $\eta$ exchanges \cite{ DAmbrosio:1994fgc, Cirigliano:2011ny}.
Higher chiral orders spoil this  cancellation and unfortunately also the cleanness of the prediction  even of the sign of  $\mathcal{A}(K_L \to \gamma \gamma)$ \cite{Pich:1995qp, Gerard:2005yk}.
When $\textrm{sgn}[ \mathcal{A}(K_L \to \gamma \gamma)] = \pm \textrm{sgn}[\mathcal{A}(K_L \to(\pi^0)^{\ast} \to  \gamma \gamma)] $, we represent $+$ or $-$ in Eq.~\eqref{eq:KLmumu:SM}. The choice of $+$ $(-)$ gives a destructive (constructive) interference between short- and long-distance contributions to $\mathcal{B}(K_L \to \mu^+ \mu^-)$ in the SM  \cite{Pich:1995qp, Gerard:2005yk} .

On the other hand, experimental results are \cite{Olive:2016xmw}
\begin{align}
\mathcal{B}(K_L \to \mu^+ \mu^-)_{\rm exp} = \left( 6.84 \pm 0.11 \right) \times10^{-9},
\label{eq:KLexp}
\end{align}
and the 90\,\% C.L. upper bound is \cite{Aaij:2017tia}
\begin{align}
\mathcal{B}(K_S \to \mu^+ \mu^-)_{\rm exp} < 0.8 \times10^{-9}.
\end{align}
Although a current bound of $\mathcal{B}(K_S \to \mu^+ \mu^-)$ is weaker than the SM prediction by 2 orders of magnitude, 
an upgrade  of the LHCb experiment is aiming to reach the SM sensitivity, specifically, the LHC Run\,$3$ (from 2021) \cite{LHCbupgrade}.
Note that the branching  ratios into the electron mode are suppressed by $m_{e}^2 / m_{\mu}^2$, and the detector sensitivity to the electron mode in the LHCb is weaker than the muonic mode.

Equations~(\ref{eq:KLmumu:SM}) and (\ref{eq:KSmumu:SM}) are predictions of pure $K_L$ and $K_S$ initial states, respectively.
In this Letter, we focus on interference between $K_L$ and $K_S$ states,
\vspace{-0.1cm}
\begin{align}
\Gamma (K \to f)_{\rm int} \propto \mathcal{A}(K_S \to f)^{\ast} \mathcal{A}(K_L \to f),
\label{eq:rough_inteference}   
\end{align}
where the initial state is the same $K^0$ (or $\overline{K}^0$), and a {\it lifetime} of this contribution is $2 \tau_S$.
Such an interference contribution is first discussed in Refs.~\cite{Bell:1990tq, Sehgal:1967wq}, 
and has been observed and utilized in many processes, e.g., $K \to \pi \pi $ \cite{Carosi:1990ms}, 
 $K \to 3 \pi^0$ \cite{Cartiglia:2003pb, Ambrosino:2005iw},  $K \to \pi^+ \pi^- \pi^0$ \cite{Batley:2005zp}, and 
 $K \to \pi^0 e^+ e^-$  \cite{Donoghue:1994yt}.

\vspace{-0.4cm}
\boldmath  
\section{
Interference between $K_L$ and $K_S$ }\unboldmath
\vspace{-0.23cm}

\begin{figure*}[t]
  \begin{center} 
   \includegraphics[width=0.42 \textwidth, 
   bb=0 0 360 238
  ]{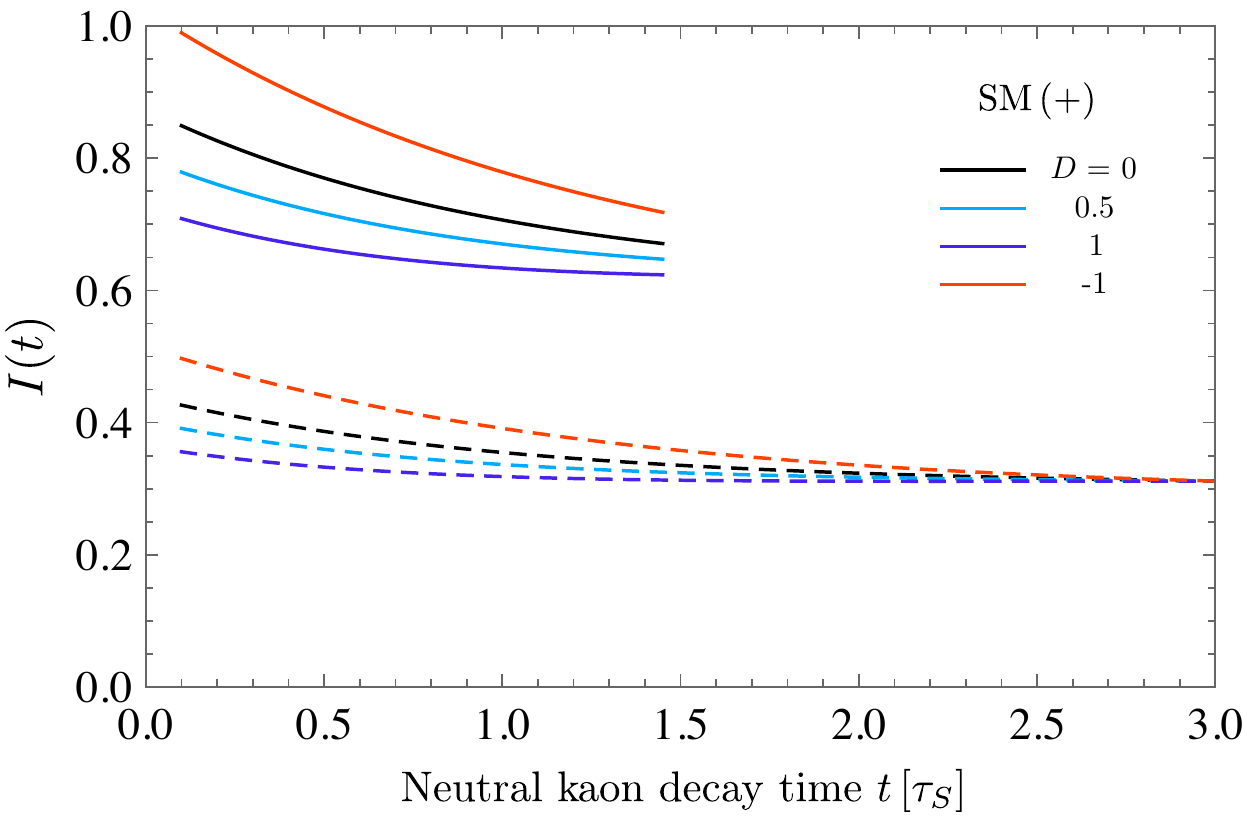}
   \hspace{1.cm}
    \includegraphics[width=0.42 \textwidth, 
    bb  = 0 0 360 238
     ]{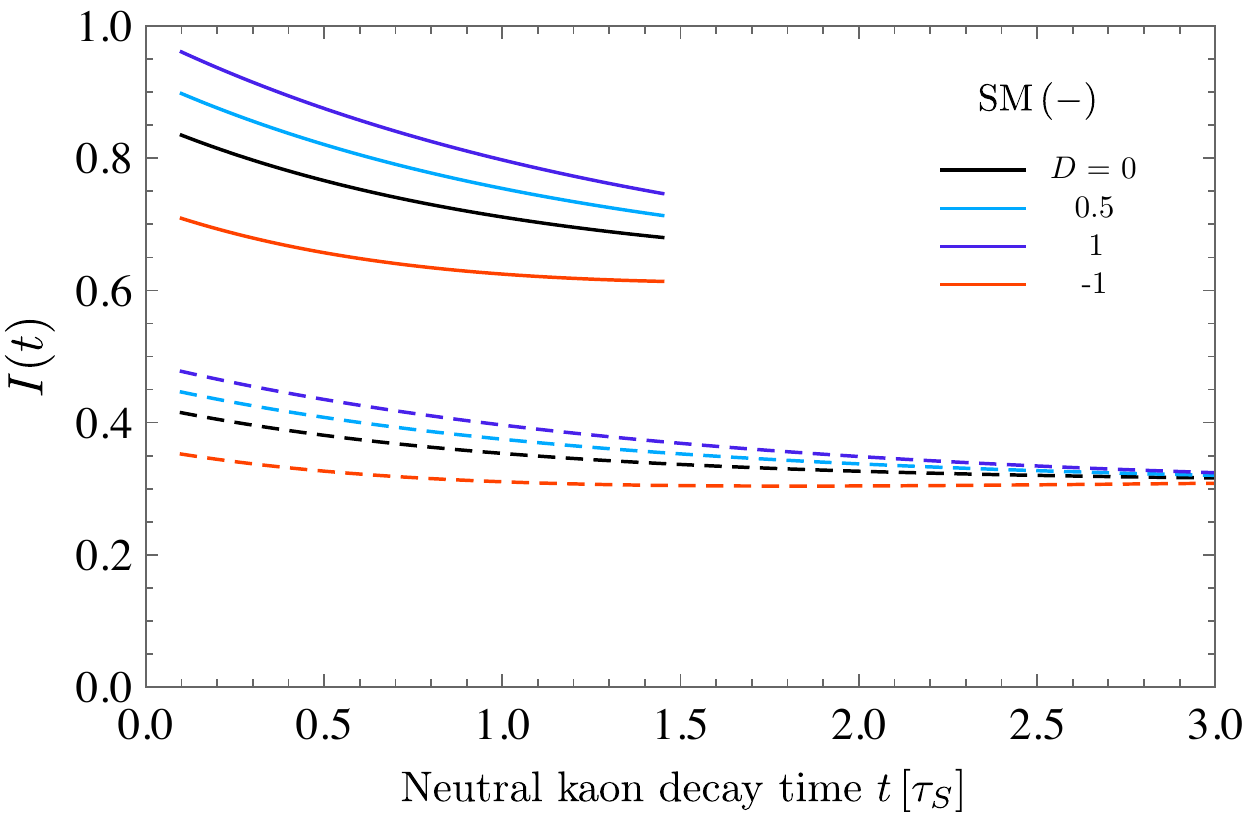}
    \vspace{-0.1cm}
    \caption{ 
   The time distributions of $K \to \mu^+ \mu^- $ [$I(t)$] are shown 
    within the SM with several choices of $D$, which are normalized by 
    the decay intensity from $0.1 \tau_S$ to $1.45 \tau_S$ (solid lines) and from $0.1 \tau_S$ to $3 \tau_S$ (dashed lines) with $D=0$.
The left and right panels correspond to the positive and negative signs of ${A}^{\mu}_{L\gamma \gamma}$ in Eq.~\eqref{eq:ALgammagamma}, respectively. }
\label{fig:time_distribution}
\end{center}
\vspace{-0.8cm}
\end{figure*}
We first review the interference contribution briefly; then, we investigate the numerical impact  in the mode of $\mu^+ \mu^-$ in the SM.
A state of $K^0$ (or $\overline{K}^0$) at $t=0 $, which is produced by, e.g., $p p \to K^0 K^- \pi^+$,   evolves into a mixture of  $K_1 $ (\textit{CP}-even) and $K_2$ (\textit{CP}-odd) states,
\vspace{-0.2cm}
\begin{align}
| \accentset{\brobor}{K}\phantom{}^{0} (t) \> 
 = & \frac{1}{ \sqrt{2} (1  \pm  \bar{\epsilon} ) } \left[ e^{- i H_S t} \left( | K_1 \>  +  \bar{\epsilon} | K_2 \> \right)  \right.\non
& \left.~~~~~~~~~~~\pm  e^{- i H_L t} \left( | K_2 \> + \bar{\epsilon} | K_1 \> \right) \right],
\end{align}
where $H_{L,S} = M_{L,S} - (i/2) \Gamma_{L,S}$, 
$
| K_{1,2} \> = (1/\sqrt{2}) ( | K^0 \>  \pm | \overline{K}^0 \> )$, and $
{CP} | K_{1,2} \> = \pm | K_{1,2} \>$.
The \textit{CP} impurity parameter $\bar{\epsilon}$ is related to $\epsilon_K$ as
$
\epsilon_K = ( \bar{\epsilon} + i  \imag A_0 / \real A_0 )/(1 + i \bar{\epsilon} \imag A_0 / \real A_0 ) 
$
with
$ \mathcal{A}(K^0 \to (\pi \pi)_{I=0}) \equiv A_0 e^{i \delta_0}$, and $\delta_0 $ is a strong phase for an $I=0$ two-pion state.

\begin{widetext}
The decay intensity of a neutral kaon beam into $f$ is 
\begin{align}
&I (t) =  \frac{1 + D}{2} \left| \< f | - \mathcal{H}^{|\Delta S|  =1}_{\rm eff} | K^0  (t)  \>\right|^2 +  \frac{1 -  D}{2} \left| \< f | - \mathcal{H}^{|\Delta S|  =1}_{\rm eff} | \overline{K}^0  (t)  \>\right|^2  \\
&=  \frac{1}{2 } \left[ \left\{  \left( 1 - 2 D \real [\bar{\epsilon} ]\right)  |\mathcal{A}(K_1)|^2 + 2 \real  \left[ \bar{\epsilon} \mathcal{A}(K_1)^{\ast} \mathcal{A}(K_2) \right]  \right\} e^{- \Gamma_S t}  + \left\{ \left( 1  - 2 D \real [\bar{\epsilon} ] \right)       |\mathcal{A}(K_2)|^2  + 2 \real  \left[ \bar{\epsilon} \mathcal{A}(K_1) \mathcal{A}(K_2)^{\ast} \right]  \right\} e^{- \Gamma_L t} \right. \non
&~~ 
+ \left\{  2  D \real  \left[ e^{-i \Delta M_K t}  \left( \mathcal{A}(K_1)^{\ast} \mathcal{A}(K_2)  + \bar{\epsilon} |\mathcal{A}(K_1) |^2 + \bar{\epsilon}^{\ast} |\mathcal{A}(K_2)|^2\right)\right]     - 4 \real [\bar{\epsilon} ]   \real  \left[ e^{-i \Delta M_K t}   \mathcal{A}(K_1)^{\ast} \mathcal{A}(K_2) \right] \right\}\left. e^{- \frac{ \Gamma_S + \Gamma_L}{2} t}\right] \non
& ~~ + \mathcal{O}( \bar{\epsilon}^2),
\label{eq:It}
\end{align}
\vspace{-0.7cm}
\end{widetext}
where $M_L - M_S \equiv \Delta M_K > 0$, $\mathcal{A}(K_{1,2}) \equiv \mathcal{A} (K_{1,2} \to f ) $, and a dilution factor $D$ is  a measure of the initial ($t=0$) asymmetry of the number of
$K^0$ and $\overline{K}^0$,
\begin{align}
 D= \frac{ K^0 - \overline{K}^0 }  { K^0 + \overline{K}^0 }.
 \label{eq:DE}
\end{align}
The term proportional to $\exp({- \Gamma_S t}) $ [or $\exp({- \Gamma_L t} )$] arises from $K_S$ (or $K_L$) decay in the mode $f$, while the term proportional to $\exp[{- (\Gamma_S + \Gamma_L) t/2}]$ represents the interference between $ K_L$ and $K_S$, whose lifetime is $2 /(\Gamma_S + \Gamma_L) \simeq 2 \tau_S$.

\vspace{-0.4cm}
\boldmath  
\section{
Interference effect on $K \to \mu^+ \mu^-$ in the SM}\unboldmath
\vspace{-0.4cm}

In the case when $f = \mu^+ \mu^-$, 
all $\mathcal{O}(\bar{\epsilon})$ terms are numerically negligible, which is certainly different situation from $K \to 2 \pi  $ and  
 $K \to 3 \pi$.
 Then, a term of Eq.~\eqref{eq:rough_inteference} is relevant, which is the first term in the second line of Eq.~\eqref{eq:It}. 
The $ |\mathcal{A}(K_{1,2})|^2$ term provides the SM prediction of $\mathcal{B}(K_{S,L} \to \mu^+ \mu^-)_{\rm SM}$ in Eqs.~(\ref{eq:KLmumu:SM}) and (\ref{eq:KSmumu:SM}) 
 \cite{Ecker:1991ru, Isidori:2003ts, Gorbahn:2006bm}, which is significantly dominated by a \textit{CP}-conserving long-distance contribution.
Within the SM, regarding the interference term, we obtain
\begin{align}
&\sum_{\rm spin} \mathcal{A}(K_1 \to \mu^+ \mu^-)^{\ast} \mathcal{A}(K_2 \to \mu^+ \mu^- )
 \non
 &   =   \frac{16 i  G_F^4 M_W^4 F_K^2  {M_{K}^2} m_{\mu}^2  \sin^2 \theta_W }{\pi^3}      \imag [\lambda_t] y'_{7A}       \non
  &~ \times \left\{  {A}^{\mu}_{L\gamma\gamma} - 2 \pi \sin^2 \theta_W \left(  \real [\lambda_t]  y'_{7A} + \real [\lambda_c]  y_c \right) \right\},
  \label{eq:SMinterference}
\end{align}
where the spin of  the muons is summed  up as $\lambda_q\equiv V^{\ast}_{qs} V_{qd}$,  
$\sin^2 \theta_W \equiv \sin^2 \hat{\theta}_W^{\overline{{\rm MS}}} (M_Z) = 0.23129(5)$  \cite{Olive:2016xmw}, 
$f_K  = \sqrt{2} F_K = 0.1556(4) \GeV$ \cite{Olive:2016xmw}, 
the top-quark contribution in next-to-leading order of QCD is
$y'_{7 A} =  - 0.654(34)$ \cite{Buchalla:1993bv, Gorbahn:2006bm} (which is defined  in  the next section),
the charm-quark contribution  in next-to-next-to-leading order of QCD is $y_c =  -2.03(32 ) \times   10^{-4} $ \cite{Gorbahn:2006bm}, and
an amplitude of the \textit{CP}-conserving long-distance contributions for $K_2$ is \cite{Isidori:2003ts, Mescia:2006jd}
\vspace{-0.2cm}
\begin{align}
  {A}^{\mu}_{L\gamma \gamma} 
  &= \frac{ \pm  2 \pi \alpha_0}{G_F^2 M_W^2 F_K {M_{K}}} \sqrt{ \frac{   \pi }{{M_{K}}}  \Gamma(K_L \to \gamma \gamma)_{\rm exp}} \non
  & \times \left(\chi_{\rm disp} + i \chi_{\rm abs} \right)\non
  &= \pm 2.01(1)\times 10^{-4} \times \left( 0.71(101)  - i \,5.21 \right),
  \label{eq:ALgammagamma}
\end{align}
with
 $ \mathcal{A} (M^2_K) = \chi_{\rm disp} + i \chi_{\rm abs}$ \cite{Mescia:2006jd}, where $\mathcal{A} (s)$  given in Refs.~\cite{GomezDumm:1998gw,Knecht:1999gb} denotes contributions from $ 2 \gamma$ intermediate state and its counterterm, and {disp} ({abs}) indicates the dispersive (absorptive) contribution,  
$\mathcal{B}(K_L \to \gamma \gamma)_{\rm exp} =  5.47(4) \times 10^{-4} $ \cite{Olive:2016xmw} and $\alpha_0 = 1/137.04$.
Here, the sign ambiguity in $ {A}^{\mu}_{L\gamma \gamma}$ comes from the unknown sign of $\mathcal{A} (K_L \to \gamma \gamma) $, and this $\pm $ corresponds to 
$\textrm{sgn}[\mathcal{A}(K_L \to \gamma \gamma)] = \pm \textrm{sgn}[\mathcal{A}(K_L \to(\pi^0)^{\ast} \to  \gamma \gamma)] $ and Eq.~\eqref{eq:KLmumu:SM}.
Obviously, the interference in Eq.~\eqref{eq:SMinterference} is proportional to the direct \textit{CP}-violating contribution.

Figure~\ref{fig:time_distribution} shows a time distribution of $K \to \mu^+ \mu^-$ in Eq.~\eqref{eq:It} with 
several choices of $D$ and the sign of ${A}^{\mu}_{L\gamma \gamma}$,
which are normalized by 
    an integrated decay intensity from $0.1 \tau_S$ to $1.45 \tau_S$ (solid lines) and to $3 \tau_S$ (dashed lines) with $D=0$.
It is shown that the interference effect emerges  prominently around $t\simeq0$, which can give $\mathcal{O}(10\%)$ difference. 
Another important point found here is that one can  probe the unknown sign  of ${A}^{\mu}_{L\gamma \gamma}$ by precise measurement of the interference correction.

\begin{widetext}
Using the result of Eq.~\eqref{eq:It}, let us define an {\it effective} branching ratio into $\mu^+ \mu^-$, which includes the interference correction and would correspond to event numbers in experiments after a removal of the $K_L$ background,
\begin{align}
&\mathcal{B}( K_S \to \mu^+ \mu^-)_{\rm eff}\non
 &= \tau_S \left[  \int_{t_{\textrm{min}}}^{t_{\textrm{max}}} dt  \left(  \Gamma(K_1)  e^{- \Gamma_S t} +   \frac{D}{8 \pi {M_{K}}} \sqrt{1 - \frac{4 m_{\mu}^2}{{M_{K}^2}}} \sum_{\rm spin}  \real \left[ e^{-i \Delta M_K t}  \mathcal{A}(K_1)^{\ast} \mathcal{A}(K_2)\right]  e^{- \frac{ \Gamma_S + \Gamma_L}{2} t} 
\right)  \varepsilon(t) \right] \non
& \times 
\left( \int_{t_\textrm{{min}}}^{t_\textrm{{max}}} dt \, e^{- \Gamma_S t}  \,\varepsilon(t)\right)^{-1},
\label{eq:BKSmumueff}
\end{align}
where $\Gamma(K_1) = \Gamma ( K_1 \to \mu^+ \mu^-)$, ${t_\textrm{{min}}}$ to  ${t_\textrm{{max}}} $ corresponds to a range of detector for $K_S$ tagging, and $ \varepsilon(t) $ is a decay-time acceptance of the detector. Note that $\mathcal{B}( K_S \to \mu^+ \mu^-)_{\rm eff} = \mathcal{B}(K_S \to \mu^+ \mu^-)_{\rm SM} $ in Eq.~\eqref{eq:KSmumu:SM}  is obtained when $D=0$ is chosen.
For the removal of the $K_L$ background, the experimental result of $K_L \to \mu^+ \mu^-$ in Eq.~\eqref{eq:KLexp}  can be utilized. 
The LHCb also measures $K_S \to \pi^+ \pi^- $ decay as a normalization mode; then, the number of produced $K_S$ can be derived. 
The number of produced $K_L$ is the same as $K_S$ to a good approximation, so that using the experimental value in Eq.~\eqref{eq:KLexp}  one can estimate and subtract the $K_L \to \mu^+ \mu^-$ background    \cite{footnote:KL}. This procedure is independent of $D$, whose dependence appears from $\mathcal{O} (\bar{\epsilon})$.
\vspace{-0.2cm}
\end{widetext}

\begin{figure*}[t]
  \begin{center} 
   \includegraphics[width=0.42 \textwidth, 
   bb=0 0 360 216
   ]{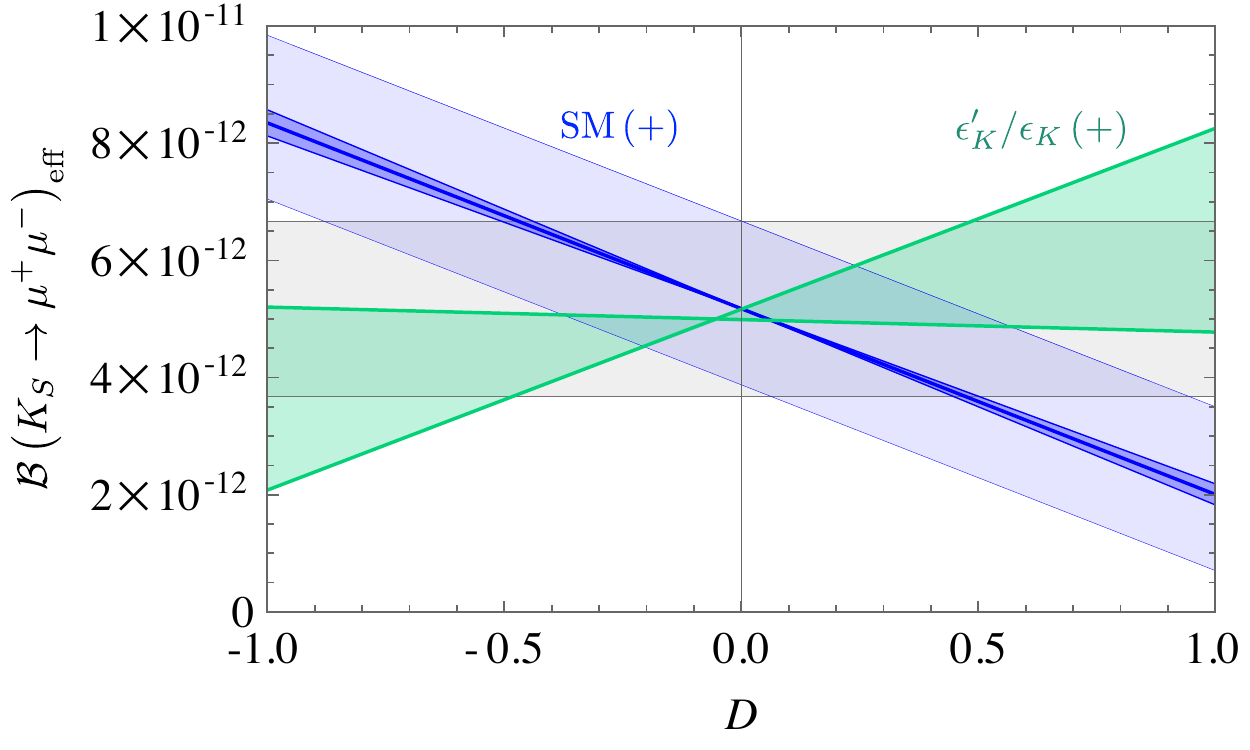}
   \hspace{1.cm}
    \includegraphics[width=0.42 \textwidth, 
    bb    = 0 0 360 216
   ]{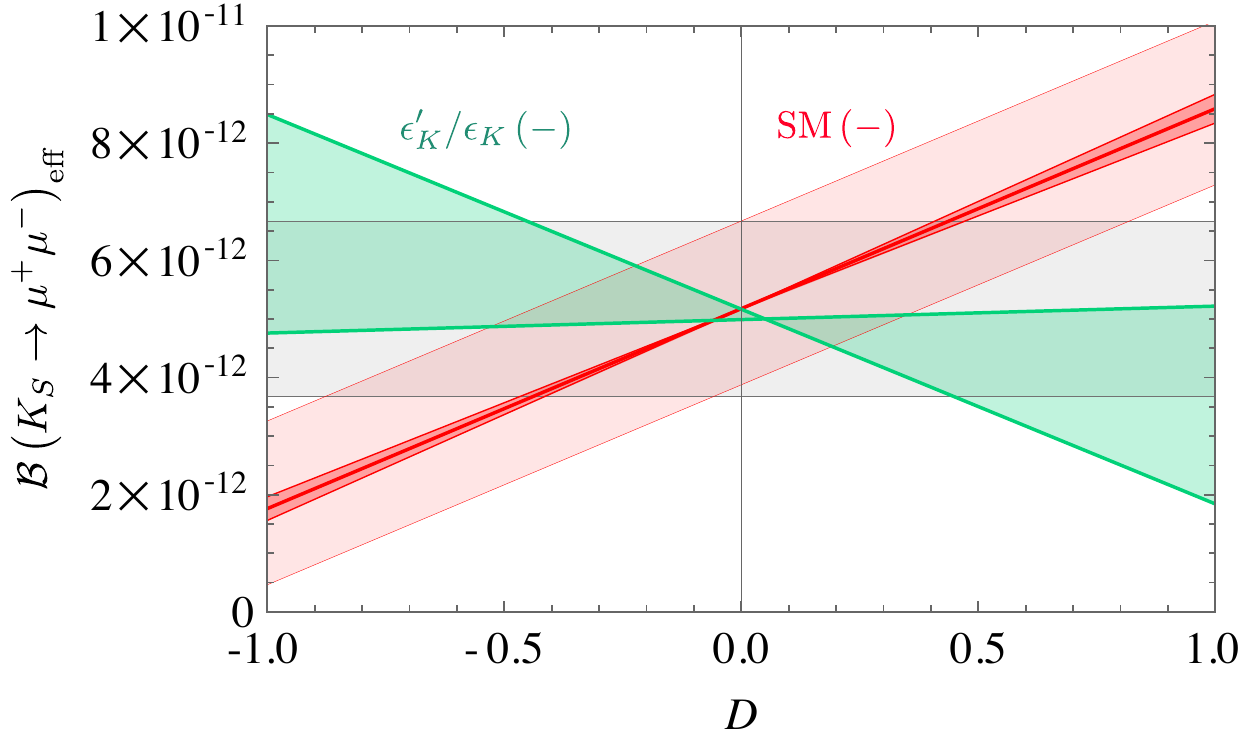}
    \vspace{-0.3cm}
    \caption{
    The effective branching ratio into $\mu^+ \mu^-$  in Eq.~\eqref{eq:BKSmumueff} as a function of the  dilution factor.
The left and right panels correspond to the positive and negative signs of ${A}^{\mu}_{L\gamma \gamma}$ in Eq.~\eqref{eq:ALgammagamma}, respectively. 
The SM predictions are represented by blue and red lines, where the darker bands stand for uncertainty from the interference in Eq.~\eqref{eq:SMinterference} and the lighter bands  denote uncertainty from $A^{\mu}_{S\gamma\gamma}$ in Eq.~\eqref{eq:BSgammagamma}.
Gray bands represent $\mathcal{B}(K_S \to \mu^+ \mu^-)_{\rm SM} $ in Eq.~\eqref{eq:KSmumu:SM}.
The $\epsilon'_K /\epsilon_K$ anomaly can be explained  at 1\,$\sigma$ in the green regions 
 within the modified $Z$-coupling model.
     }
\label{fig:BReff}
\end{center}
\vspace{-0.9cm}
\end{figure*}

We investigate the effective branching ratio in Eq.~\eqref{eq:BKSmumueff} as a function of $D$ in Fig.~\ref{fig:BReff}.
Here, the experimental setup of the LHCb detector is adopted:  
the decay-time acceptance  is 
$
\varepsilon(t) = \exp( - \beta t)
$, where  $\beta \simeq 86\,(\textrm{ns})^{ -1}$ \cite{Diego}.
The range of the detector for selecting $K \to \mu^+ \mu^-$ is $ t_{\textrm{min}} = 8.95\,\textrm{ps} = 0.1 \tau_S$ and  $ t_{\textrm{max}} = 130\,\textrm{ps} = 1.45 \tau_S$ \cite{Diego}.
Gray bands represent $\mathcal{B}(K_S \to \mu^+ \mu^-)_{\rm SM} $ in Eq.~\eqref{eq:KSmumu:SM}.
The blue and red lines are  the SM predictions, where the lighter (darker) bands stand for uncertainty from   $A^{\mu}_{S\gamma\gamma}$ [from the interference term in Eq.~\eqref{eq:SMinterference}], which is 
an amplitude of the \textit{CP}-conserving long-distance contributions for $K_1$ \cite{Ecker:1991ru, Isidori:2003ts, Mescia:2006jd},
\vspace{-0.2cm}
\begin{align}
 {A}^{\mu}_{S \gamma \gamma}  
 & = \frac{\pi \alpha_0}{ G_F^2 M_W^2 F_K {M_{K}} |H(0)|} \sqrt{ \frac{\pi}{{M_{K}} } \Gamma(K_S \to \gamma \gamma)_{\rm exp}} \non
 & ~~~\times  \left( \mathcal{I}_{\rm disp} + i \mathcal{I}_{\rm abs} \right) \non
 & = 2.48  (35) \times 10^{-4} \times ( -2.83 + i \,1.22),
 \label{eq:BSgammagamma}
\end{align}
with 
 $\mathcal{I} ( m_{\mu}^2 / M_{K}^2 , m_{\pi^{\pm}}^2 / M_{K}^2) = \mathcal{I}_{\rm disp} + i \mathcal{I}_{\rm abs} $, where a two-loop function $\mathcal{I} (a,b)$ from the $2\pi^{\pm} 2 \gamma$  intermediate state  is given in Refs.~\cite{Isidori:2004rb, Ecker:1991ru},  
$
\mathcal{B}(K_S \to \gamma \gamma)_{\rm exp} =  2.63 (17) \times 10^{-6}
$  \cite{Olive:2016xmw}
and the pion one-loop function  $H(0) = 0.331+ 
 i 0.583$  \cite{Ecker:1991ru} are used.
Since this evaluation includes a 17\% enhancement of the amplitude by a final-state interaction of the pions and it is reasonable for on-shell  but not off-shell photon emission, a 30\% uncertainty to the branching ratio is taken \cite{Isidori:2003ts}.

It is found that the interference affects the branching ratio at $ \mathcal{O}(60\%)$  and  the unknown sign  of ${A}^{\mu}_{L\gamma \gamma}$   can be uncovered  if $D = \mathcal{O}(1)$ can be used.
Note that the error of $ {A}^{\mu}_{S \gamma \gamma} $ dominates the uncertainties of all lines. Since the dispersive treatment \cite{Colangelo:2016ruc}
will sharpen  ${A}^{\mu}_{S \gamma \gamma}  $,  the interference and $\mathcal{B}(K_S \to \mu^+ \mu^-)_{\rm SM}$ will, hence,  be transparent in these figures.
{We also comment on  the difference between two effective branching ratios with different dilution factors $D$ and $D'$,  $\mathcal{B}( K_S \to \mu^+ \mu^-)_{\rm eff} (D) - \mathcal{B}( K_S \to \mu^+ \mu^-)_{\rm eff} (D') \propto (D - D') $, where $D>0$ and  $D'<0$  for $K^0$ and $\overline{K}^0$  tagging, respectively.  This measurement does not receive a large uncertainty from $ {A}^{\mu}_{S \gamma \gamma} $, so that one can more clearly determine the sign of $\mathcal{A}(K_L \to \gamma \gamma)$. }

Note that because $\sigma(p p\to K^0 X ) \simeq \sigma(p p\to \overline{K}^0 X ) $,  $D $ would be $0$ as a standard of the LHCb experiment. 
We propose two methods for generating $K^0$--$\overline{K}^0$ asymmetry in the neutral kaon signals.
The first one is the tagging of a charged kaon  which accompanies the neutral kaon beam. An $\mathcal{O}(30\%)$ of prompt $K^0$  accompanies $K^{-}$ through  $p p \to K^0 K^{-} X  $  \cite{Diego}.
Such a charged kaon track with a $K\to \mu^+ \mu^-$ signal can be tagged by using the RICH detectors.
This charged kaon tagging has been utilized to tag $B^0_s$ in the LHCb experiment \cite{LHCb:2011aa}.
A similar tagging would be possible for $\Lambda^0$ through $p p \to K^0 \Lambda^0 X  $ with $\Lambda^0 \to  p \pi^-$ \cite{Adinolfi:2012qfa}.
Another proposal is a charged-pion tagging using $ p p \to K^{\ast +} X  \to K^0 \pi^+ X$. A similar charged-pion  tagging for $D^0$ ($D^{\ast +} \to D^0 \pi^+$) has been achieved in the LHCb  experiment \cite{Aaij:2011in}.

\vspace{-0.5cm}
\boldmath  
\section{
Probing New Physics}\unboldmath
\vspace{-0.4cm}

We now investigate the influence of new physics on the interference.
In general  new physics, only three operators can contribute to $K \to \mu^+ \mu^-$.  Then,  the interference term in Eq.~\eqref{eq:SMinterference} can be extended to
\vspace{-0.1cm}
\begin{align}
&\sum_{\rm spin} \mathcal{A}(K_1 \to \mu^+ \mu^-)^{\ast} \mathcal{A}(K_2 \to \mu^+ \mu^- )
 \non
 &   =\frac{8 G_F^4 M_W^4 F_K^2  {M_{K}^2} m_{\mu}^2 }{\pi^4}  \non
  & \times \left[   \left( 1 - \frac{ 4 m_{\mu}^2}{{M_{K}^2}} \right)\left\{ \left({A}^{\mu}_{S \gamma \gamma}\right)^{\ast} + \frac{{M_{K}^2}}{M_W^2}\real {\tilde{y}'_S} \right\}    i  \frac{{M_{K}^2}}{M_W^2}\imag {\tilde{y}'_S} \right.\non
 &
 +  \left\{ 2 i \pi \sin^2 \theta_W\left( \imag \left[ \lambda_t \right]y'_{7A}  +\imag  \tilde{y}'_{7A} \right)  - i \frac{{M_{K}^2}}{M_W^2} \imag {\tilde{y}'_P} \right\}  \non
  &~~~~ \times \Big\{   - 2 \pi \sin^2 \theta_W \left(  \real\left[ \lambda_t \right] y'_{7A} + \real\left[\lambda_c \right] y_c +\real \tilde{y}'_{7A}     \right) \non
  &  ~~~~\left. \left. + {A}^{\mu}_{L\gamma\gamma} +  \frac{{M_{K}^2}}{M_W^2}\real {\tilde{y}'_P} \right\}  \right],
  \label{eq:NPinterference}
\end{align}
where the Wilson coefficients are defined in \cite{Mescia:2006jd}
\vspace{-0.cm}
\begin{eqnarray}
\mathcal{H}^{|\Delta S|  =1}_{\rm eff}
=  \frac{G_F^2 m_s m_{\mu}  }{\pi^2}  \Big\{ \tilde{y}'_S \left( \overline{s} \gamma_5 d \right) \left( \overline{\mu}  \mu\right)  +\tilde{y}'_P \left( \overline{s} \gamma_5 d \right) \left( \overline{\mu} \gamma_5 \mu\right) \Big\}
\non
 + \frac{G_F \alpha }{\sqrt{2}} \left( \lambda_t y'_{7A}  + \tilde{y}'_{7A} \right) 
\left( \overline{s} \gamma_{\mu} \gamma_5 d \right) \left( \overline{\mu} \gamma^{\mu} \gamma_5 \mu\right) + {\rm H.c.}~~~~~~
\end{eqnarray}
Here, new physics contributions are represented by $\tilde{y}'$, and 
$\alpha \equiv \alpha_{\overline{{\rm MS}}}(M_Z) =1/127.95$ \cite{Olive:2016xmw}.
We find that the interference in Eq.~\eqref{eq:NPinterference} is still a genuine direct \textit{CP}-violating contribution.
The new physics contributions ($\tilde{y}'_S $, $\tilde{y}'_P$, and $\tilde{y}'_{7A} $) to $\Gamma(K_{1,2} \to \mu^+ \mu^- ) $ are given in Ref.~\cite{Mescia:2006jd}.

The following is a specific example of new physics: we focus on a modified $Z$-coupling model \cite{Buras:2014sba, Buras:2015yca, Buras:2015jaq, Endo:2016tnu, Bobeth:2017xry}, which can easily explain a 2.8$\sigma$--2.9$\sigma$ discrepancy in  $\epsilon'_K / \epsilon_K$  between the measured values and the predicted one at next-to-leading order 
 \cite{Bai:2015nea,Buras:2015yba, Kitahara:2016nld}.
 In this model, after the electroweak symmetry breaking, the following flavor-changing $Z$ interactions emerge:
 \begin{align}
\mathcal{H}^{|\Delta S|  =1}_{\rm eff} = - \Delta_L^{\rm NP} \overline{s} \gamma_{\mu} P_L d Z^{\mu} + \left(L \leftrightarrow R \right) + {\rm H.c.}
 \end{align}
In our analysis, we assume that the new physics is only left handed and that it is pure imaginary, for simplicity: $\Delta_R^{\rm NP} = \real \Delta_L^{\rm NP} = 0$. 
According to Ref.~\cite{Endo:2016tnu}, the $\epsilon'_K / \epsilon_K$ discrepancy is explained at the 1\,$\sigma$ level by the range of $-1.05\times 10^{-6} < \imag \Delta_L^{\rm NP} < -0.50\times 10^{-6}$ without conflict with $\epsilon_K$ and 
$\mathcal{B} (K_L \to \mu^+ \mu^-)$.
 This range corresponds to
 \begin{align}
0.86 \times 10^{-4} < \imag \tilde{y}'_{7A} < 1.82 \times 10^{-4}, ~~\tilde{y}'_S  = \tilde{y}'_P = 0.
 \end{align}

Green bands in Fig.~\ref{fig:BReff} show that the effective branching ratio into $\mu^+ \mu^-$  in Eq.~\eqref{eq:BKSmumueff}, which can explain the $\epsilon'_K / \epsilon_K$ discrepancy at 1\,$\sigma$.
It is observed that the interference vanishes or flips the sign compared to the SM predictions. This is because the interference is proportional to the direct \textit{CP} violation ($ \imag\left[\lambda_t \right]y'_{7A}  +\imag  \tilde{y}'_{7A} $) and 
$\imag[\lambda_t] y'_{7 A} = -0.92\times 10^{-4}$.

The other new physics scenario that can explain   the $\epsilon'_K / \epsilon_K$ discrepancy \cite{Kitahara:2016otd} will be presented in a forthcoming article \cite{Diego2}.

\vspace{-0.4cm}
\section{Discussion And Conclusions}

In this Letter, we have studied the interference between $K_L$ and $K_S$ in $K \to \mu^+ \mu^-$ within the SM and the modified $Z$-coupling model, which could be probed by the future upgrade of  the LHCb experiment.
We  have pointed out that the interference is a genuine direct \textit{CP}-violation effect,  so that one can investigate   direct \textit{CP} violation  by precise  measurement of $K \to \mu^+ \mu^-$.
It is found that within the SM the interference can amplify the effective branching ratio of $K_S \to \mu^+ \mu^-$  in Eq.~\eqref{eq:BKSmumueff} by $\mathcal{O}(60\%)$  by determining  the unknown sign of $\mathcal{A}(K_L \to \gamma \gamma)$, which can much reduce the theoretical uncertainty of $\mathcal{B}(K_L \to \mu^+ \mu^-)$. 
  It is also shown that in the  modified $Z$-coupling model, accounting for  the $ \epsilon'_K /\epsilon_K$ anomaly,  the interference is predicted to vanish or flip the sign.

 Such an investigation of  the direct \textit{CP} violation of kaon decay is  important for, of course, $\epsilon'_K / \epsilon_K$ tension and also as  a cross-check of the KOTO experiment, which is probing a \textit{CP}-violating $K_L \to \pi^0 \nu \overline{\nu}$ decay and will reach SM sensitivity in 2021 \cite{KOTO, KOTO2}.
 
A similar study would be possible for 
$K_S \to \pi^+ \pi^- \pi^0$ using the interference in the LHCb experiment.
Although there are significant background  events from  $K_L \to \pi^+ \pi^- \pi^0$, 
 the Dalitz analysis of the momenta of the three pions can cut the background
 \cite{Batley:2005zp,DAmbrosio:1994vba}.

\vspace{-.4cm}
\section*{Acknowledgments}

We would like to thank Diego Martinez Santos for valuable comments about the LHCb experiment and Ulrich Nierste for reading the manuscript. 
T.~K. is also grateful to Yuto Minami for fruitful discussions.
G. D. was supported in part by MIUR under Project No. 2015P5SBHT (PRIN 2015) and by the INFN research initiative ENP.

\providecommand{\href}[2]{#2}
\begingroup\raggedright


\begin{thebibliography}{99}
\bibitem{footnote:K+loop}
The charged-kaon loop contribution is suppressed by $1 - M_{K^+}^2 / M_{K^0}^2$.
\bibitem{footnote:KL}
Under the setup of the LHCb detector we obtain $\int dt \Gamma (K_L \to \mu^+ \mu^-)_{\rm exp} e^{-\Gamma_L t} \varepsilon(t)/ [\int dt \Gamma (K_1)_{\rm SM}  $ $ e^{-\Gamma_S t} \varepsilon(t)]  = 2.9. $
\end{thebibliography}
\end{document}